\documentclass[pss]{wiley2sp} % provides pss two-column style
\usepackage{amsmath}
\newcommand{\be}{\begin{equation}}
\newcommand{\ee}{\end{equation}}
\newcommand{\bea}{\begin{eqnarray}}
\newcommand{\eea}{\end{eqnarray}}
\begin{document}

% Title of the article
\title{Exact results for a generalized spin-1/2 Ising--Heisenberg diamond chain
with the second-neighbor interaction between nodal spins}

% Abbreviated title for the page headers
\titlerunning{Generalized spin-1/2 Ising--Heisenberg diamond chain}

% Authors
\author{%
  Bohdan Lisnyi\textsuperscript{\textsf{\bfseries 1,\bfseries 2}},
  Jozef Stre\v{c}ka\textsuperscript{\textsf{\bfseries 2},\Ast}}

% Abbreviated list of authors for the page headers
\authorrunning{B. Lisnyi and J. Stre\v{c}ka}

%E-mail-address of corresponding author
\mail{e-mail \textsf{jozef.strecka@upjs.sk}, Phone: +421-55-2342276, Fax: +421-55-6222124}

% author's affiliations/addresses
\institute{%
  \textsuperscript{1}\,Institute for Condensed Matter Physics, National Academy of Sciences of Ukraine, 1 Svientsitskii Street, Lviv 79011, Ukraine \\
  \textsuperscript{2}\,Insitute of Physics, Faculty of Science, P. J. \v{S}af\'{a}rik University, Park Angelinum 9, 040 01 Ko\v{s}ice, Slovak Republic }

\received{XXXX, revised XXXX, accepted XXXX} % do not change, will be filled in by the publisher
\published{XXXX} % do not change, will be filled in by the publisher

% Please select about four verbal keywords for your manuscript.

\abstract{The ground state and thermodynamics of a generalized spin-1/2 Ising--Heisenberg diamond chain
with the second-neighbor interaction between nodal spins are calculated exactly using the mapping method
based on the decoration-iteration transformation. Rigorous results for the magnetization, susceptibility,
and heat capacity are investigated in dependence on temperature and magnetic field for the frustrated
diamond spin chain with the antiferromagnetic Ising and Heisenberg interactions. It is demonstrated that the second-neighbor
interaction between nodal spins gives rise to a greater diversity of low-temperature magnetization curves,
which may include an intermediate plateau at two-third of the saturation magnetization related to the classical
ferrimagnetic (up-up-up-down-up-up-...) ground state with translationally broken symmetry besides an intermediate one-third
magnetization plateau reflecting the translationally invariant quantum ferrimagnetic (monomer-dimer) spin arrangement.}

\maketitle

\section{Introduction}

Decorated spin chains, which can be exactly treated by combining the decoration-iteration transformation with the transfer-matrix method \cite{s5,s6,roja09,spla10,rojas11,bell13},
are of practical importance for the qualitative interpretation of magnetic phenomena emergent in real solid-state materials.
In particular, several exactly solved Ising--Heisenberg decorated spin chains provide in-depth understanding of a striking interplay between
geometric spin frustration and quantum fluctuations, which may manifest itself through various intriguing phenomena such as the appearance
of intermediate plateaux in low-temperature magnetization curves, the formation of additional maxima in the temperature dependence of the
susceptibility and specific heat, the enhanced magnetocaloric effect during the adiabatic demagnetization and so on. Despite a certain
oversimplification, some exactly solved Ising--Heisenberg spin chains afford a plausible quantitative description of the magnetic behavior of real
spin-chain materials \cite{s0,s3,exp10,sah12,str12,han13,oha13}.

During the past decade, the natural mineral azurite Cu$_3$(CO$_3$)$_2$(OH)$_2$ has attracted a considerable research interest, because it provides
a long sought experimental realization of the spin-1/2 diamond chain with spectacular magnetic properties \cite{kik04,ki05l,ki05ptp,rul08}.
Owing to this fact, a lot of attention has been paid to a rigorous treatment of various versions of the Ising--Heisenberg diamond chain \cite{dos2,can09,roj11,lis3,ana12,roj12,ana13,gal13,gal14,anan14}. Although a correct description of magnetic properties of the azurite would require modeling based on a more complex spin-1/2 Heisenberg model \cite{jes11,hon11}, the simplified but still exactly tractable spin-1/2 Ising--Heisenberg diamond chain qualitatively reproduces the most prominent experimental features reported for the azurite such as an intermediate one-third magnetization plateau as well as the double-peak temperature dependences of specific heat and susceptibility \cite{kik04,ki05l,ki05ptp,rul08}.
In addition, it has been theoretically predicted that the asymmetric spin-1/2 Ising--Heisenberg diamond chain \cite{lis3} relevant to the azurite may display much richer magnetic behavior than its symmetric counterpart \cite{dos2}, since the asymmetry in exchange interactions along the diamond sides may cause new peculiarities such as an existence of the additional magnetization plateau
at zero magnetization and/or unusual temperature dependence of zero-field specific heat with three distinct round peaks \cite{lis3}.

To provide consistent description of all experimental data reported so far for the azurite (i.e. low-temperature magnetization curve, INS and NMR data, magnetic susceptibility and specific heat \cite{kik04,ki05l,ki05ptp,rul08}), the first-principles calculations based on the density functional theory have been recently combined with the state-of-the-art numerical calculations \cite{jes11,hon11}. This powerful combination of accurate methods has convincingly evidenced that the asymmetric spin-1/2 Heisenberg diamond chain suggested in the early studies as a feasible model of the azurite \cite{kik04,ki05l,ki05ptp,rul08} must be inevitably extended in order to include the second-neighbor interaction between the nodal spins as well as non-negligible
inter-chain interactions \cite{jes11}. However, a closer inspection of the exchange interactions in the azurite allows one to map a rather complex three-dimensional Heisenberg model
to the asymmetric spin-1/2 Heisenberg diamond chain with the second-neighbor interaction between the nodal spins under additional small refinement of all exchange constants \cite{hon11}.

Bearing all this in mind, the main purpose of this work is to examine magnetic properties of the asymmetric spin-1/2 Ising--Heisenberg diamond chain refined by
the second-neighbor interaction between the nodal spins. It will be evidenced that the second-neighbor interaction between the nodal spins
may cause an emergence of the additional intermediate plateau at two-thirds of the saturation magnetization, the nature of which is completely the same as recently proposed
for the symmetric spin-1/2 Ising--Heisenberg diamond chain accounting for an extra four-spin interaction \cite{gal13,gal14}. It is worthwhile to remark, moreover,
that an investigation of the generalized spin-1/2 Ising--Heisenberg diamond chain with the second-neighbor interaction between the nodal spins is intriguing because
this model is isomorphic with the spin-1/2 Ising--Heisenberg doubly decorated chain \cite{dos1,lis1} refined by the additional further-neighbor interactions
as well as the spin-1/2 Ising--Heisenberg tetrahedral chain \cite{valjpcm08}. Consequently, the exact results presented hereafter for the generalized
spin-1/2 Ising--Heisenberg diamond chain with the second-neighbor interaction between the nodal spins have some important implications also for the remarkable quantum
antiferromagnetic order experimentally observed in the copper-based tetrahedral chain Cu$_3$Mo$_2$O$_9$ \cite{hase08,hama08,kuro11,hase11,mats12,kuro14}.

The organization of this paper is as follows. Section \ref{model} provides an exact solution of the generalized spin-1/2 Ising--Heisenberg diamond chain with the second-neighbor
Ising interaction between the nodal spins and the XYZ Heisenberg interaction between the interstitial spins in the spirit of the decoration-iteration mapping transformation. The most interesting results obtained for the ground-state phase diagrams are discussed together with typical magnetic field and temperature dependences of the magnetization, susceptibility, and specific heat
in Section \ref{result}. A few experimental results reported previously for solid-state representatives of the diamond spin chain Cu$_3$(CO$_3$)$_2$(OH)$_2$ and the tetrahedral spin chain Cu$_3$Mo$_2$O$_9$ are also briefly qualitatively interpreted with the help of the studied model in Section \ref{experiment}. Finally, some conclusions and future outlooks are drawn in Section \ref{conclusion}.

\section{Ising--Heisenberg diamond chain}
\label{model}

\begin{figure}[t]
\begin{center}
\includegraphics[width=0.95\columnwidth]{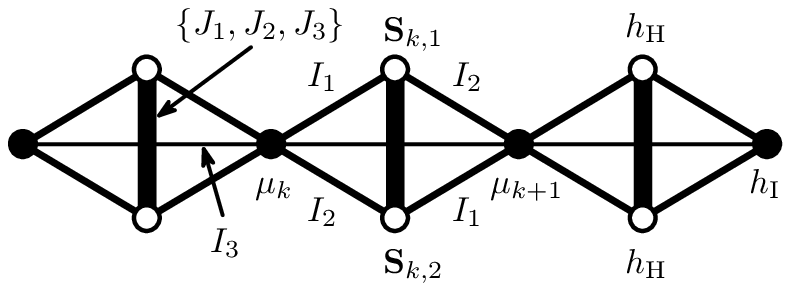}
\end{center}
\vspace{-0.5cm}
\caption{A fragment of the generalized spin-1/2 Ising--Heisenberg diamond chain. The Ising spins $\mu_{k}$ and $\mu_{k+1}$ at two nodal sites and the Heisenberg spins $\mathbf{S}_{k,1}$ and $\mathbf{S}_{k,2}$ at two interstitial sites of the $k$-th primitive cell are marked.}
\label{fig1}
\end{figure}

Let us consider the generalized spin-1/2 Ising--Heisenberg diamond chain in a presence of the external magnetic field. A primitive cell of the diamond spin chain is determined by the nodes $k$ and $k+1$ as illustrated in Fig.~\ref{fig1}. All nodal sites are occupied by the Ising spins $\mu_{k}=\pm 1/2$, which are coupled with their neighbors solely through the Ising interactions. On the contrary, two interstitial sites $(k,1)$ and $(k,2)$ from the $k$-th primitive cell are occupied by the Heisenberg spins ${\mathbf{S}}_{k,1}$ and ${\mathbf{S}}_{k,2}$, which are mutually coupled through the spatially anisotropic XYZ Heisenberg interaction. For further convenience, the total Hamiltonian of the generalized spin-1/2 Ising--Heisenberg diamond chain can be defined as a sum over cell Hamiltonians $\hat{\cal H}_k$:
\begin{eqnarray}
\hat {\cal H} = \sum\limits_{k=1}^N \hat {\cal H}_k,
\label{Ht}
\end{eqnarray}
where each cell Hamiltonian $\hat{\cal H}_k$ involves all the interaction terms belonging to the $k$-th primitive cell
\begin{eqnarray}
\hat {\cal H}_k \!& {=} &\!
J_1 \hat S^x_{k,1}\hat S^x_{k,2} {+} J_2 \hat S^y_{k,1}\hat S^y_{k,2} {+} J_3 \hat S^z_{k,1}\hat S^z_{k,2}
+ I_3 \mu_{k} \mu_{k{+}1}
\nonumber \\
&& {} {+}  \mu_{k} \left(I_1 \hat S^z_{k,1} {+} I_2 \hat S^z_{k,2} \right)
      {+}  \mu_{k+1} \left(I_2 \hat S^z_{k,1} {+} I_1 \hat S^z_{k,2} \right)
\nonumber \\
&& {} - h_{\rm{H}} \left(\hat S^z_{k,1} + \hat S^z_{k,2} \right)
- \frac{h_{\rm{I}}}{2} \left(\mu_{k} + \mu_{k+1} \right),
\label{Hk}
\end{eqnarray}
In above, $\hat{S}^\alpha_{k,i}$ ($\alpha = x, y, z$; $i= 1, 2$) denote spatial components of the spin-1/2 operator, the parameters $J_1$, $J_2$ and $J_3$ determine the spatially anisotropic XYZ interaction between the nearest-neighbor Heisenberg spins, $I_1$ and $I_2$ label the Ising interactions between the nearest-neighbor Ising and Heisenberg spins residing the diamond sides, $I_3$ represents the second-neighbor Ising interaction between the nodal spins, $h_{\rm{I}}$ and $h_{\rm{H}}$ are the magnetic fields acting on the Ising and Heisenberg spins, respectively. It should be mentioned that a few particular cases of the Hamiltonian (\ref{Hk}) have been extensively studied in the past. The particular case $I_2=I_3=0$ (or $I_1=I_3=0$) corresponds to the spin-1/2 Ising--Heisenberg doubly decorated chain \cite{dos1,lis1}, the other particular case $I_1 = I_2$ and $I_3=0$ corresponds to the symmetric spin-1/2 Ising--Heisenberg diamond chain \cite{dos2}, while the most symmetric special case $I_1 = I_2 = I_3$ corresponds to the spin-1/2 Ising--Heisenberg tetrahedral chain \cite{valjpcm08}.

Let us calculate the partition function of the generalized spin-1/2 Ising--Heisenberg diamond chain. With regard to a validity of commutative relation between different cell Hamiltonians $[\hat{\cal H}_i, \hat{\cal H}_j] = 0$, the partition function ${\cal Z}$ can be partially factorized into the following product:
\begin{equation}
{\cal Z} = \sum_{\{\mu_k\}} \prod_{k=1}^N \mbox{Tr}_{\mathbf{S}_{k,1},\mathbf{S}_{k,2}} \exp \left(-\beta \hat {\cal H}_k \right),
\label{Z1}
\end{equation}
where $\beta=1/(k_{\rm{B}}T)$, $k_{\rm{B}}$ is the Boltzmann's constant, $T$ is the absolute temperature, the symbol $\sum_{\{\mu_k\}}$ denotes summation over a complete set of spin states of the Ising spins and the symbol $\mbox{Tr}_{\mathbf{S}_{k,1},\mathbf{S}_{k,2}}$ marks a trace over spin degrees of freedom of two Heisenberg spins from the $k$-th primitive cell. To proceed further with a calculation, one necessarily needs to evaluate the effective Boltzmann's factor
\begin{equation}
{\cal Z}_k(\mu_{k}, \mu_{k+1}) = \mbox{Tr}_{\mathbf{S}_{k,1},\mathbf{S}_{k,2}} \exp \left(-\beta \hat {\cal H}_k \right),
\label{BF}
\end{equation}
which naturally appears behind the product symbol in the factorized form (\ref{Z1}) of the partition function. For this purpose, it is quite advisable to pass to the matrix representation of the
cell Hamiltonian $\hat {\cal H}_k$ in the basis spanned over four available states of two Heisenberg spins $S^z_{k,1}$ and $S^z_{k,2}$:
\begin{eqnarray}
|\!\uparrow, \uparrow \rangle_{k} &=& \left |\uparrow \right \rangle_{k,1} \left |\uparrow \right \rangle_{k,2},
\quad
|\! \downarrow, \downarrow \rangle_{k} = \left |\downarrow \right \rangle_{k,1} \left |\downarrow \right \rangle_{k,2},
\nonumber \\
|\! \uparrow, \downarrow \rangle_{k} &=& \left |\uparrow \right \rangle_{k,1} \left |\downarrow \right \rangle_{k,2},
\quad
|\!\downarrow, \uparrow \rangle_{k} = \left |\downarrow \right \rangle_{k,1} \left |\uparrow \right \rangle_{k,2},
\label{B}
\end{eqnarray}
whereas $|\!\!\uparrow\rangle_{k,i}$ and $|\!\!\downarrow\rangle_{k,i}$ denote two eigenvectors of the spin operator $\hat{S}^z_{k,i}$ with the respective eigenvalues $S^z_{k,i} = 1/2$ and $-1/2$. After a straightforward diagonalization of the cell Hamiltonian $\hat {\cal H}_k$ one obtains the following four eigenvalues:
\begin{eqnarray}
{\cal E}_{1,2}  &=&
- \frac{h_{\rm{I}}}{2} \left(\mu_{k} + \mu_{k+1} \right) + I_3 \mu_{k} \mu_{k+1} + \frac{J_3}{4} \pm Q_{+} ~,
\nonumber \\
{\cal E}_{3,4} & = &
- \frac{h_{\rm{I}}}{2} \left(\mu_{k} + \mu_{k+1} \right) + I_3 \mu_{k} \mu_{k+1} - \frac{J_3}{4} \pm Q_{-} ~,
\label{Ek}
\end{eqnarray}
where
\[
Q_{\pm}\! = \!\frac{1}{2} \sqrt{\!\left(\!\!\frac{J_1 \!\mp\! J_2}{2}\!\!\right)^2 \!\!\! +
\left [(I_1 \!\pm\! I_2)(\mu_{k} \!\pm\! \mu_{k+1}) - h_{\rm{H}} \!\mp\! h_{\rm{H}} \right ]^2} .
\]
Now, one may simply use the eigenvalues (\ref{Ek}) in order to calculate the Boltzmann's factor (\ref{BF}) according to the relation ${\cal Z}_k (\mu_{k}, \mu_{k+1}) = \sum\limits_{i=1}^4 {\rm e}^{-\beta {\cal E}_i }$. The explicit form of the relevant Boltzmann's factor can be subsequently replaced through the generalized decoration-iteration transformation \cite{s5,s6,roja09,spla10}:
\begin{eqnarray}
&\!\! {\cal Z}_k (\mu_{k}, \mu_{k+1})\!=\! 2\exp \! \left[\frac{\beta h_{\rm{I}}}{2} \left(\mu_{k} + \mu_{k+1} \right) - \beta I_3 \mu_{k} \mu_{k+1} \right]
\nonumber \\
& {} \times \left [ \exp \! \left(\!-\frac{\beta J_3}{4}\! \right) \cosh \left(\beta Q_{+}\right) + \exp \! \left(\!\frac{\beta J_3}{4}\!\right) \cosh \left(\beta Q_{-}\right) \right ]
\nonumber \\
& {} = A \exp \!\left[\beta R \mu_{k} \mu_{k+1} + \frac {\beta H_0}{2} \left(\mu_{k} + \mu_{k+1} \right) \right]\!.
\label{DIT}
\end{eqnarray}
The mapping parameters $A$, $R$, and $H_0$ are unambiguously determined by a 'self-consistency' condition of the decoration-iteration transformation (\ref{DIT}), which requires a validity of the mapping transformation independently of the spin states of two Ising spins $\mu_{k}$ and $\mu_{k+1}$ involved therein. The decoration-iteration mapping transformation (\ref{DIT}) satisfies the 'self-consistency' condition if and only if
\begin{eqnarray}
A &=&\left [{\cal Z}_k \left( \frac12, \frac12 \right) {\cal Z}_k \left(-\frac12, -\frac12 \right)
{\cal Z}_k^2 \left( \frac12, -\frac12 \right)\right ]^{{1}/{4}},
\nonumber \\
\beta R &=& \ln \left[ \frac{{\cal Z}_k \left( \frac12, \frac12 \right) {\cal Z}_k \left(-\frac12, -\frac12 \right)}{
{\cal Z}_k^{2} \left( \frac12, -\frac12 \right)} \right],
\nonumber \\
\beta H_0 &=& \ln \left[ \frac{{\cal Z}_k \left( \frac12, \frac12 \right)}{{\cal Z}_k \left(-\frac12, -\frac12\right)} \right].
\label{MP}
\end{eqnarray}
An important consequence directly follows from the explicit formulas for the mapping parameters (\ref{MP}), which imply that the effective coupling $\beta R$ is the only mapping parameter
that depends on the second-neighbor interaction $I_3$ between the nodal spins through a trivial shift $\beta R = \beta R |_{I_3=0} - \beta I_3$ while the other two mapping parameters
$A$ and $\beta H_0$ are totally independent of $I_3$.

By inserting the decoration-iteration transformation (\ref{DIT}) into Eq. (\ref{Z1}) one readily gets a rigorous mapping relation between the partition function ${\cal Z}$ of the generalized spin-1/2 Ising--Heisenberg diamond chain and the partition function ${\cal Z}_0$ of the simple spin-1/2 Ising chain with the effective nearest-neighbor interaction $R$ and the effective magnetic field $H_0$:
\begin{equation}
{\cal Z} = A^N {\cal Z}_0 (\beta, R, H_0).
\label{MR}
\end{equation}
It is worthy to recall that the spin-1/2 Ising chain in a magnetic field is well-known exactly tractable model, which can be solved for instance through the classical transfer-matrix method \cite{s2}. The partition function of the spin-1/2 Ising chain in a magnetic field is determined by a sum of two eigenvalues of the transfer matrix ${\cal Z}_0 = \lambda_{+}^N + \lambda_{-}^N$, which are for the sake of completeness given by the following expressions:
\begin{eqnarray}
\lambda_{\pm} = {\rm e}^{\frac{\beta R}{4}} \left[ \cosh \left(\frac{\beta H_0}{2}\right) \pm \sqrt{\sinh^2 \left(\frac{\beta H_0}{2}\right) + {\rm e}^{-\beta R}} \right]\!.
\nonumber
\end{eqnarray}
From this point of view, the exact calculation of the partition function ${\cal Z}$ of the generalized spin-1/2 Ising--Heisenberg diamond chain with the second-neighbor interaction between the nodal spins is formally completed.

Exact results for other thermodynamic quantities follow quite straightforwardly from the mapping relation (\ref{MR}) between the partition functions. In the thermodynamic limit $N \to \infty$, the free energy per unit cell can be evaluated from the formula
\begin{equation}
f = \lim_{N \to \infty} - \frac{1}{N} \beta^{-1} \ln {\cal Z} = -\frac{1}{\beta} (\ln A + \ln \lambda_{+}),
\label{FE}
\end{equation}
which also allows a straightforward calculation of the entropy $s$ and the heat capacity $c$ per unit cell:
\begin{equation}
s = k_{\rm{B}} \beta^2  \frac{\partial f}{\partial \beta}~,
\qquad
c = - \beta \frac{\partial s}{\partial \beta}~.
\label{SC}
\end{equation}
The single-site magnetization of the Ising spins $m_{\rm{I}} \equiv \langle \mu_{k} + \mu_{k+1} \rangle /2$ and the single-site magnetization of the Heisenberg spins $m_{\rm{H}} \equiv \left \langle \hat S_{k,1}^z + \hat S_{k,2}^z \right \rangle/2$ can be obtained by a differentiation of the free energy (\ref{FE}) with respect to the particular magnetic fields:
\begin{equation}
m_{\rm{I}}= - \frac{\partial f}{\partial h_{\rm{I}}}~,
\qquad
m_{\rm{H}}= - \frac{1}{2} \frac{\partial f}{\partial h_{\rm{H}}}~.
\label{MIH}
\end{equation}
Once evaluated, the total magnetization can be expressed in terms of both single-site magnetizations through the formula $ m=(m_{\rm{I}} + 2m_{\rm{H}})/3$.

\section{Results and Discussion}
\label{result}

In this section, let us discuss the most interesting results obtained for the generalized spin-1/2 Ising--Heisenberg diamond chain with the second-neighbor interaction between the nodal spins. For simplicity, our further attention will be focused just on the particular case with the antiferromagnetic Ising interactions $I_1, I_2, I_3>0$ and the antiferromagnetic XXZ Heisenberg interaction  $J_{1} = J_{2} = J \Delta$, $J_{3} = J > 0$. The main motivation for a detailed study of the antiferromagnetic spin-1/2 Ising--Heisenberg diamond chain lies in that this special case should exhibit the most obvious manifestations of a mutual interplay between the geometric spin frustration and local quantum fluctuations. Note furthermore that the dimensionless parameter $\Delta$ determines a spatial anisotropy in the XXZ Heisenberg interaction and the special case of $\Delta = 1$ corresponds to the isotropic Heisenberg coupling between the nearest-neighboring interstitial spins. To further reduce the number of free interaction parameters, we will also assume equal magnetic fields acting on the Ising and Heisenberg spins $h=h_{\rm{I}}=h_{\rm{H}}$ what physically corresponds to setting g-factors of the Ising and Heisenberg spins equal one to each other. Another important observation can be made from the Hamiltonian (\ref{Hk}) of the generalized spin-1/2 Ising--Heisenberg diamond chain that is invariant with respect to the inter-change $I_1 \longleftrightarrow I_2$ under simultaneous re-numbering of the interstitial sites $k,1 \longleftrightarrow k,2$ and hence, one may also consider $I_1 \geq I_2$ without loss of generality. This fact allows us to introduce a difference between two Ising interactions along the diamond sides $\delta I = I_1 - I_2 \geq 0$ and to use the stronger among two Ising interactions as the energy unit when defining the following set of dimensionless interaction parameters:
\begin{equation}
\tilde{J}=\frac{J}{I_1},  \quad \delta \tilde{I}=\frac{\delta I}{I_1}, \quad \tilde{I_3}=\frac{I_3}{I_1}, \quad \tilde{h}=\frac{h}{I_1}.
\label{DP}
\end{equation}
The reduced interaction parameters given by Eq. (\ref{DP}) measure a relative strength of the Heisenberg interaction, the asymmetry of two Ising interactions along the diamond sides, the second-neighbor interaction between the nodal spins, and the external magnetic field, all normalized with respect to the stronger Ising interaction ($I_1$) along the diamond sides. It is quite evident that the accessible values of the parameter $\delta \tilde{I}$, whose physical sense lies in the degree of asymmetry of two Ising interactions along the diamond sides, are then restricted to the interval $\delta \tilde{I} \in [0,1]$.

First, let us examine the ground state of the generalized spin-1/2 Ising--Heisenberg diamond chain. The ground state can be trivially connected to the lowest-energy eigenstate of the cell Hamiltonian (\ref{Ek}) obtained by taking into account all four states of two nodal Ising spins $\mu_{k}$ and $\mu_{k+1}$ that enter into the respective eigenvalues. Depending on a mutual competition between the interaction parameters $\tilde{J}$, $\Delta$, $\delta \tilde{I}$, $\tilde{I}_3$ and $\tilde{h}$ one finds in total five different ground states: the saturated paramagnetic state SPA, two classical ferrimagnetic states FRI$_1$ and FRI$_2$, the quantum ferrimagnetic state QFI and the quantum antiferromagnetic state QAF given by the eigenvectors 
\begin{eqnarray}
|\mbox{SPA} \rangle \! &{=}& \prod\limits_{k=1}^N| + \rangle_k ~|\!\uparrow, \uparrow \rangle_{k},
\nonumber\\
|\mbox{FRI}_1 \rangle \! &{=}& \prod\limits_{k=1}^N| - \rangle_k ~|\!\uparrow, \uparrow \rangle_{k},
\nonumber\\
|\mbox{QFI} \rangle \! &{=}& \prod\limits_{k=1}^N | + \rangle_k
~\frac{1}{\sqrt{2}} \big[|\!\uparrow, \downarrow \rangle_{k} - |\!\downarrow, \uparrow \rangle_{k} \big],
\nonumber\\
|\mbox{QAF} \rangle \! &{=}& \!\!\left\{\begin{array}{l}
\prod\limits_{k=1}^N \left|[-]^k \right\rangle_k
\!\left[A_{[-]^k} |\!\uparrow, \downarrow \rangle_{k} - A_{[-]^{k{+}1}} |\!\downarrow, \uparrow \rangle_{k} \right]
\\[4mm]
\prod\limits_{k=1}^N \!\left|[-]^{k{+}1} \!\right\rangle_k
\!\!\left[A_{[-]^{k{+}1}} |\!\uparrow, \downarrow \rangle_{k} {-} A_{[-]^{k}} |\!\downarrow, \uparrow \rangle_{k} \right]
\end{array} \right.\!,
\nonumber\\
|\mbox{FRI}_2\!\rangle \! &{=}& \!\!\left\{\begin{array}{l}
\prod\limits_{k=1}^N \left|[-]^k \right\rangle_k ~|\!\uparrow, \uparrow \rangle_{k}
\\[4mm]
\prod\limits_{k=1}^N \left|[-]^{k+1} \right\rangle_k ~|\!\uparrow, \uparrow \rangle_{k}
\end{array} \right. .
\label{GS}
\end{eqnarray}
In above, the ket vector $|\pm\rangle_k$ determines the state of the nodal Ising spin $\mu_k = \pm 1/2$, the symbol $[-]^k \in \{-,+\}$ marks the sign of the number $(-1)^k$, the spin states relevant to two Heisenberg spins from the $k$th primitive cell are determined by the notation (\ref{B}) and the probability amplitudes $A_{\pm}$ are explicitly given by the expressions:
\begin{equation}
A_{\pm}=\frac{1}{\sqrt{2}}\sqrt{1 \mp \frac{\delta \tilde{I}}{\sqrt{(\delta \tilde{I})^2 + (\tilde{J} \Delta)^2}}}.
\label{PA}
\end{equation}
The eigenenergies per primitive cell that correspond to the respective ground states (\ref{GS}) are given as follows:
\begin{eqnarray}
\tilde{\cal E}_{\rm{SPA}} &{=}& \frac{\tilde{J}}{4} + 1 - \frac{\delta \tilde{I}}{2} + \frac{\tilde{I}_3}{4}
- \frac{3\tilde{h}}{2},
\nonumber\\
\tilde{\cal E}_{\rm{FRI}_1} &{=}& \frac{\tilde{J}}{4} - 1 + \frac{\delta \tilde{I}}{2} + \frac{\tilde{I}_3}{4}
- \frac{\tilde{h}}{2},
\nonumber\\
\tilde{\cal E}_{\rm{QFI}} &{=}& - \frac{\tilde{J}}{4} - \frac{\tilde{J}\Delta}{2} + \frac{\tilde{I}_3}{4}
- \frac{\tilde{h}}{2},
\nonumber\\
\tilde{\cal E}_{\rm{QAF}} &{=}& - \frac{\tilde{J}}{4} - \frac{1}{2}\sqrt{(\delta \tilde{I})^2 + (\tilde{J} \Delta)^2}
- \frac{\tilde{I}_3}{4},
\nonumber\\
~\tilde{\cal E}_{\rm{FRI}_2}\!\! &{=}& \frac{\tilde{J}}{4} - \frac{\tilde{I}_3}{4} - \tilde{h}.
\label{EE}
\end{eqnarray}

Let us shortly comment on respective spin arrangement inherent to the ground states (\ref{GS}). At high magnetic fields, the antiferromagnetic spin-1/2 Ising--Heisenberg diamond chain naturally ends up at SPA ground state with all nodal Ising and interstitial Heisenberg spins fully polarized by the external magnetic field. Contrary to this, the ground-state spin alignment is  
much more diverse at lower magnetic fields when either one of three ferrimagnetic ground states (FRI$_1$, FRI$_2$ or QFI) or the unique quantum antiferromagnetic ground state QAF is realized.
The ground state FRI$_1$ corresponds to a classical ferrimagnetic spin arrangement, in which all interstitial Heisenberg spins are fully aligned with the magnetic field and all nodal Ising spins point in an opposite direction due to the antiferromagnetic coupling with their nearest-neighbor interstitial spins. However, it is energetically more favorable for the Heisenberg spin pairs to form the singlet-dimer state provided that the antiferromagnetic coupling between the Heisenberg spins is strong enough. Under this condition, the antiferromagnetic spin-1/2 Ising--Heisenberg diamond chain rests in the quantum ferrimagnetic ground state QFI with a character of the dimer-monomer state, because the nodal Ising spins tend to align with the magnetic field on behalf of a spin frustration  that effectively switches off the coupling between the nearest-neighbor interstitial and nodal spins. The second-neighbor coupling between the nodal Ising spins may additionally cause the antiferromagnetic alignment of the nodal Ising spins at low enough magnetic fields, which consequently leads to the unique quantum antiferromagnetic ground state QAF. The most striking feature of QAF is that the antiferromagnetic alignment of the nodal Ising spins is surprisingly transferred to a quantum superposition of two intrinsically antiferromagnetic states ($|\!\! \uparrow, \downarrow \rangle_{k}$ and $|\!\! \downarrow, \uparrow \rangle_{k}$) of the Heisenberg spin pairs, which fall into a perfect singlet-dimer state just for the symmetric diamond chain $\delta \tilde{I} = 0$ while any asymmetry $\delta \tilde{I} \neq 0$ causes according to Eqs.~(\ref{GS})-(\ref{PA}) the spin-singlet-like state with a non-zero staggered magnetization on the Heisenberg spin pairs. The second-neighbor interaction between the nodal Ising spins may be also responsible for an appearance of another classical ferrimagnetic ground state FRI$_2$ with translationally broken symmetry, which cannot be in principle found in the spin-1/2 Ising--Heisenberg diamond chain without this interaction term \cite{lis3}. The ground state FRI$_2$ can be characterized by a full alignment of all interstitial Heisenberg spins with the magnetic field, whereas the antiferromagnetic arrangement of the nodal Ising spins arises from the antiferromagnetic second-neighbor coupling $\tilde{I}_3$ in between them. 
 
\begin{figure*}
\begin{center}
\includegraphics[width=\textwidth]{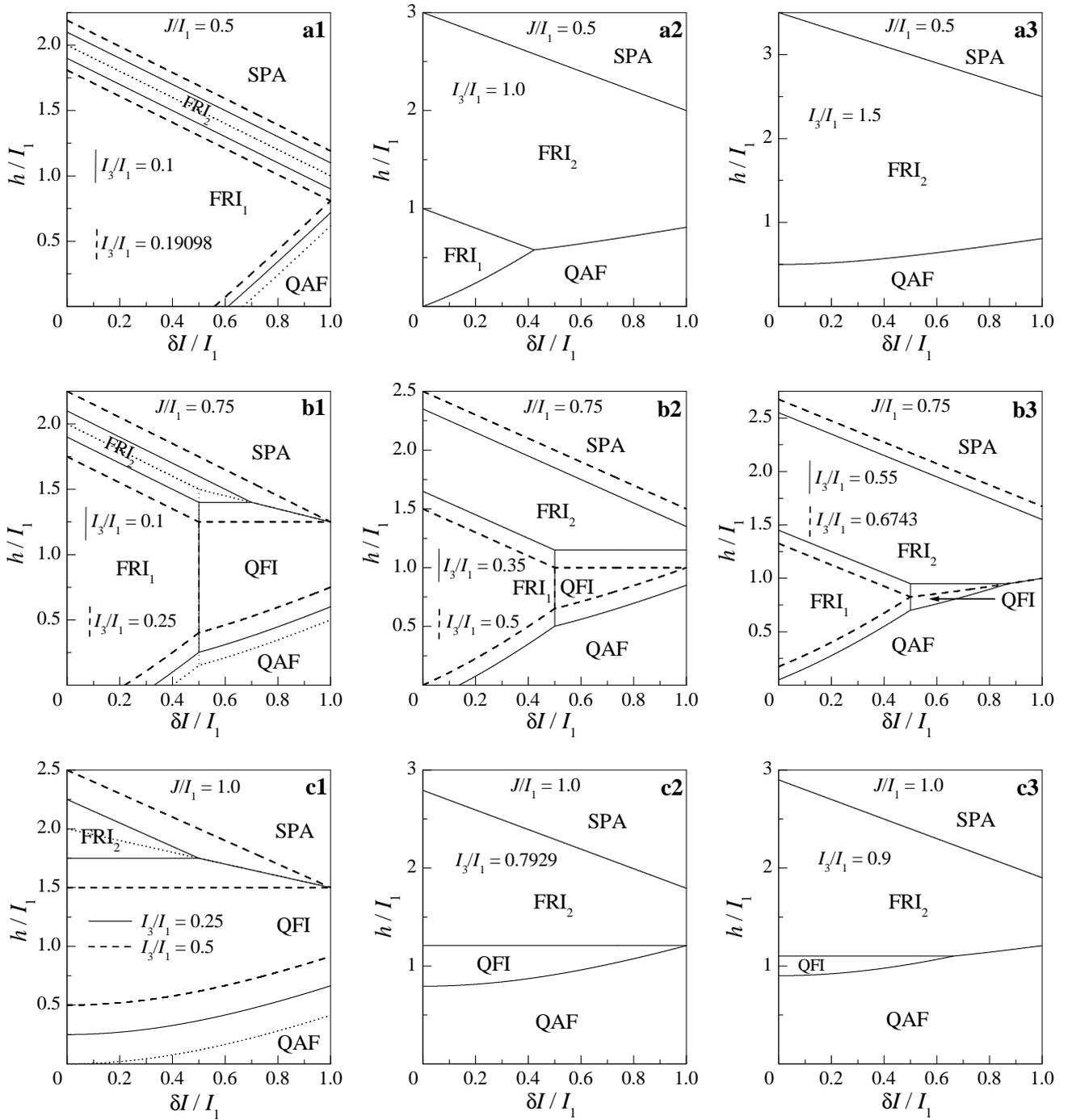}
\end{center}
\caption{Ground-state phase diagrams in the $\delta\tilde{I} - \tilde{h}$ plane constructed by considering several values of the relative strength of the second-neighbor interaction $\tilde{I}_3$ and three different relative strengths of the isotropic Heisenberg interaction ($\Delta=1$): (a) $\tilde{J}=0.5$, (b) $\tilde{J}=0.75$, and (c) $\tilde{J}=1.0$. The dotted lines shown in Fig.~\ref{fig2}(a1),(b1),(c1) correspond to the special case without the second-neighbor coupling $\tilde{I}_3=0$.}
\label{fig2}
\end{figure*}
 
Now, let us proceed to a detailed analysis of the ground-state phase diagram. The ground-state phase diagram in the $\delta \tilde{I}-\tilde{h}$ plane in an absence of the second-neighbor interaction between the nodal spins might have three different topologies depending on a size of the parameter $\tilde{J}(1+\Delta)$ \cite{lis3}: the topology of type 1 for $\tilde{J}(1+\Delta)\leq1$ 
shown in Fig.~\ref{fig2}(a1), the topology of type 2 for $1< \tilde{J}(1+\Delta) <2$ displayed in Fig.~\ref{fig2}(b1), and the topology of type 3 for $\tilde{J}(1+\Delta)\geq2$ illustrated in Fig.~\ref{fig2}(c1). The relevant ground-state boundaries for the special case of $\tilde{I}_3=0$ are shown in Fig.~\ref{fig2}(a1),(b1),(c1) by dotted lines for the illustrative case of the isotropic Heisenberg interaction ($\Delta = 1$). In what follows, we will concentrate our attention only to the influence of the second-neighbor interaction $\tilde{I}_3$ on the topology of the respective ground-state phase diagrams, whereas the reader interested in more details concerned with the special case $\tilde{I}_3=0$ is referred to Ref. \cite{lis3}. 

Consider first the changes in the ground-state phase diagram of type 1 invoked by the strengthening of the second-neighbor interaction $\tilde{I}_3$. It is quite obvious from Fig.~\ref{fig2}(a1) that 
the direct field-induced transition between the FRI$_1$ and SPA phases observable for the special case $\tilde{I}_3=0$ along the line $\tilde{h} = 2 - \delta \tilde{I}$ vanishes on account of a presence of the band-like region pertinent to the FRI$_2$ phase. A cross-section of the band-like region in parallel to the field axis $\tilde{h}$ equals to $2\tilde{I}_3$, which means that the field range inherent to the FRI$_2$ phase becomes the greater the stronger the second-neighbor interaction  $\tilde{I}_3$ is. At zero magnetic field, the FRI$_1$ phase is replaced with the QAF phase above the boundary value  
\begin{eqnarray}
\delta \tilde{I} = \frac{2 - \tilde{J} - \tilde{I}_3}{2} - \frac{(\tilde{J} \Delta)^2}{2\left(2 - \tilde{J} - \tilde{I}_3 \right)},
\label{PB1}
\end{eqnarray}
which monotonically decreases with increasing the second-neighbor interaction $\tilde{I}_3$ until it reaches zero at the threshold value $\tilde{I}_3 = 2 - \tilde{J}(1+\Delta)$ [see Fig.~\ref{fig2}(a2)]. If the strength of the second-neighbor interaction $\tilde{I}_3$ is from the interval
\begin{eqnarray}
1 - \frac{1}{2} \! \left(\tilde{J} + \sqrt{(\tilde{J}\Delta)^2 + 1}\right) \leq \tilde{I}_3 \leq 2 - \frac{1}{2} \tilde{J}(1+\Delta),
\label{PB2}
\end{eqnarray}
then, the ground-state phase diagram contains a special triple point with the coordinates:  
\begin{eqnarray}
\delta \tilde{I} \!&=&\! \frac{1}{3} \!\! \left[2\left(4 - 2\tilde{I}_3 - \tilde{J} \right) \!-\! \sqrt{3(\tilde{J}\Delta)^2 \!+\! \left(4 - 2\tilde{I}_3 - \tilde{J}\right)^2} \right]\!\!, \nonumber\\
\tilde{h} \!&=&\! \frac{1}{3} \!\! \left[2 \left(\tilde{J} - 1 \right) + \tilde{I}_3 \!+\! \sqrt{3 (\tilde{J}\Delta)^2 \!+\! \left(4 - 2\tilde{I}_3 - \tilde{J} \right)^2} \right]\!\!,
\nonumber\\
\label{TP1}
\end{eqnarray}
at which the FRI$_1$, QAF and FRI$_2$ phases coexist together [see Fig.~\ref{fig2}(a2)]. The coexistence point of the FRI$_1$, QAF and FRI$_2$ phases gradually moves towards lower values of the asymmetry parameter $\delta \tilde{I}$ with increasing the second-neighbor interaction $\tilde{I}_3$ (along the imaginary part of transition line between the QAF and FRI$_2$ phases) until it reaches the symmetric point $\delta \tilde{I}=0$ for $\tilde{I}_3 = 2 - \tilde{J}(1+\Delta)/2$. Herewith the FRI$_1$ phase completely disappears from the ground-state phase diagram as it is illustrated in Fig.~\ref{fig2}(a3), whereas a further increase in the second-neighbor interaction $\tilde{I}_3$ only extends the area pertinent to the FRI$_2$ phase but it does not qualitatively change the topology of the phase diagram.

The effect of the second-neighbor interaction $\tilde{I}_3$ upon the ground-state phase diagram of type 2 is quite similar as in the previous case, but the relevant phase diagram is in general much more complicated due to a presence of the QFI phase residing a parameter space with a rather high asymmetry of two Ising interactions along the diamond sides. The second-neighbor interaction $\tilde{I}_3$ repeatedly gives rise to the band-like region corresponding to the FRI$_2$ phase, which emerges instead of the direct field-induced transition between the FRI$_1$ and SPA phases unlike the special case $\tilde{I}_3=0$ [see Fig.~\ref{fig2}(b1)]. In addition, the parameter region inherent to the FRI$_2$ phase wedges in between the SPA and QFI phases, whereas the apex of this wedge forms the triple point that determines a coexistence of the SPA, QFI and FRI$_2$ phases at [Fig.~\ref{fig2}(b1)] 
\begin{eqnarray}
\delta \tilde{I} \!&=&\! 2 - \tilde{J}(1+\Delta) + 2 \tilde{I}_3, \nonumber \\
\tilde{h} \!&=&\! \tilde{J}(1+\Delta) - \tilde{I}_3.
\label{TP2}
\end{eqnarray}
This triple point is shifted towards higher values of the asymmetry parameter $\delta \tilde{I}$ with increasing of the second-neighbor interaction $\tilde{I}_3$ until it completely vanishes from the phase diagram for $\tilde{I}_3 > [\tilde{J}(1+\Delta) - 1]/2$ [see Fig.~\ref{fig2}(b2)]. Besides, two phase boundaries between the FRI$_2$-QFI and QFI-QAF phases are gradually approaching each other  upon further increase of the second-neighbor coupling $\tilde{I}_3$ until both transition lines meet at a new triple point whenever
\begin{eqnarray}
\tilde{I}_3 \geq \tilde{J}(1+\Delta) - \frac{1}{2} \left(\tilde{J} + \sqrt{\left(\tilde{J}\Delta \right)^2 + 1} \right).
\label{CTP3}
\end{eqnarray}
Apparently, the aforementioned triple point defines a coexistence of the FRI$_2$, QFI and QAF phases given by
\begin{eqnarray}
\delta \tilde{I} &=& \sqrt{ \left(\tilde{J} + 2 \tilde{J}\Delta - 2 \tilde{I}_3 \right)^2 - (\tilde{J}\Delta)^2}, \nonumber\\
\tilde{h} &=& \tilde{J}(1+\Delta) - \tilde{I}_3,
\label{TP3}
\end{eqnarray}
which can be clearly seen in Fig.~\ref{fig2}(b2)-(b3). The locus of the last triple point moves to lower values of the asymmetry parameter $\delta \tilde{I}$ with increasing the second-neighbor interaction $\tilde{I}_3$ (along the imaginary part of the transition line between the FRI$_2$ and QAF phases), which consequently reduces the parameter region inherent to the QFI phase [see Fig.~\ref{fig2}(b3)]. If the second-neighbor interaction equals to
\begin{eqnarray}
\tilde{I}_3 = \tilde{J} \left(\frac{1}{2}+\Delta \right) - \frac{1}{2} \sqrt{\left(\tilde{J}\Delta \right)^2 + \left[\tilde{J}(1+\Delta) - 2\right]^2},
\label{CTP4}
\end{eqnarray}
then, all three aforedescribed triple points determining a coexistence of the FRI$_1$-FRI$_2$-QFI, FRI$_1$-QFI-QAF, and FRI$_2$-QFI-QAF phases merge together owing to a complete disappearance of the QFI phase from the ground-state phase diagram as displayed in Fig.~\ref{fig2}(b3). As a result, the phase diagram gains for stronger values of the second-neighbor interaction $\tilde{I}_3$ the same topology as described previously in Fig.~\ref{fig2}(a2) with only one triple point determining a phase coexistence between the FRI$_1$, FRI$_2$ and QAF phases [cf. Fig.~\ref{fig2}(b2) with Fig.~\ref{fig2}(a2)]. If the second-neighbor interaction exceeds the threshold value $\tilde{I}_3 > 2 - \tilde{J}(1 + \Delta)/2$, the triple point corresponding to the phase coexistence between the FRI$_1$, FRI$_2$ and QAF phases vanishes and one recovers qualitatively the same phase diagram as illustrated in Fig.~\ref{fig2}(a3). 

Last, let us comment on changes in the ground-state phase diagram of type 3 caused by the second-neighbor interaction as shown in the lower panel of Fig.~\ref{fig2}. The most fundamental difference is that the FRI$_2$ phase does not instantaneously appear in the relevant ground-state phase diagram upon rising the second-neighbor interaction $\tilde{I}_3$ from zero in contrast to the previous two cases. Indeed, the FRI$_2$ phase emerges first in between the QFI and SPA phases just if the second-neighbor interaction is stronger than the boundary value $\tilde{I}_3 \geq [\tilde{J}(1+\Delta) - 2]/2$. The parameter region inherent to the FRI$_2$ phase is then delimited by the symmetric point $\delta \tilde{I} = 0$ and the triple point (\ref{TP2}) determining a coexistence of the SPA, QFI and FRI$_2$ phases. The triple point of the phase coexistence SPA-QFI-FRI$_2$ is shifted towards higher values of the asymmetry parameter $\delta \tilde{I}$ upon strengthening of the second-neighbor interaction $\tilde{I}_3$ until it completely vanishes from the phase diagram for $\tilde{I}_3 > [\tilde{J}(1+\Delta) - 1]/2$ [see Fig.~\ref{fig2}(c1)]. The next triple point determining a coexistence of the FRI$_2$, QFI and QAF phases occurs whenever the second-neighbor coupling satisfies the condition (\ref{CTP3}), whereas the locus of this triple point given by Eq. (\ref{TP3}) gradually moves towards lower values of the asymmetry parameter $\delta \tilde{I}$ (along the imaginary part of the transition line between the FRI$_2$ and QAF phases) as the second-neighbor interaction further strengthens [see Fig.~\ref{fig2}(c2)-(c3)]. The aforedescribed triple point cannot be found in the ground-state phase diagram for $\tilde{I}_3 > \tilde{J}(1+\Delta)/2$ due to a complete disappearance of the QFI phase and the phase diagram finally recovers the same topology as discussed previously for Fig.~\ref{fig2}(a3).      

\begin{figure*}[t!]
\begin{center}
\includegraphics[width=\textwidth]{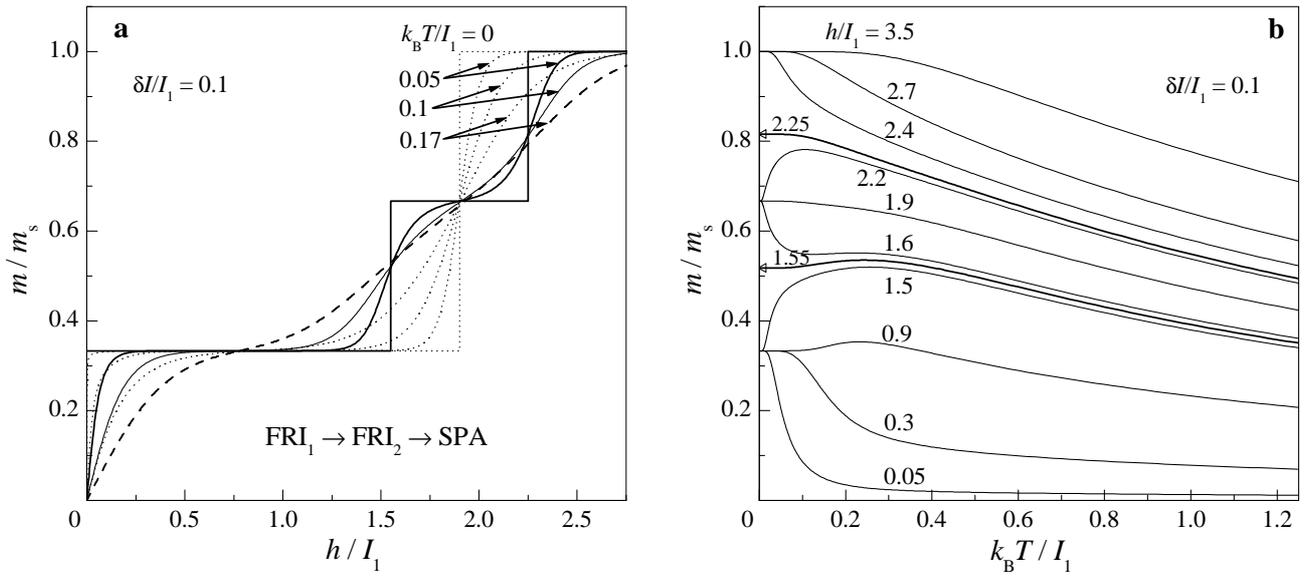}
\end{center}
\vskip-3mm
\caption{(a) The total magnetization as a function of the magnetic field at a few different temperatures for the particular case of the Heisenberg interaction $\Delta=1$ and $\tilde{J}=0.75$ by assuming one fixed value of the asymmetry parameter $\delta \tilde{I}=0.1$. The dotted lines correspond to the special case without the second-neighbor interaction $\tilde{I}_3=0$, the solid lines  to the particular case with the second-neighbor interaction $\tilde{I}_3=0.35$ of a moderate strength. (b) Thermal variations of the total magnetization for the particular case of the Heisenberg interaction $\Delta=1$ and $\tilde{J}=0.75$, the asymmetry parameter $\delta \tilde{I}=0.1$ and the second-neighbor interaction $\tilde{I}_3=0.35$ at several values of the magnetic field. The triangle symbols denote critical fields at which two different ground states coexist together.}
\vskip10mm
\label{fig3}
\end{figure*}

\begin{figure*}[h!]
\begin{center}
\includegraphics[width=\textwidth]{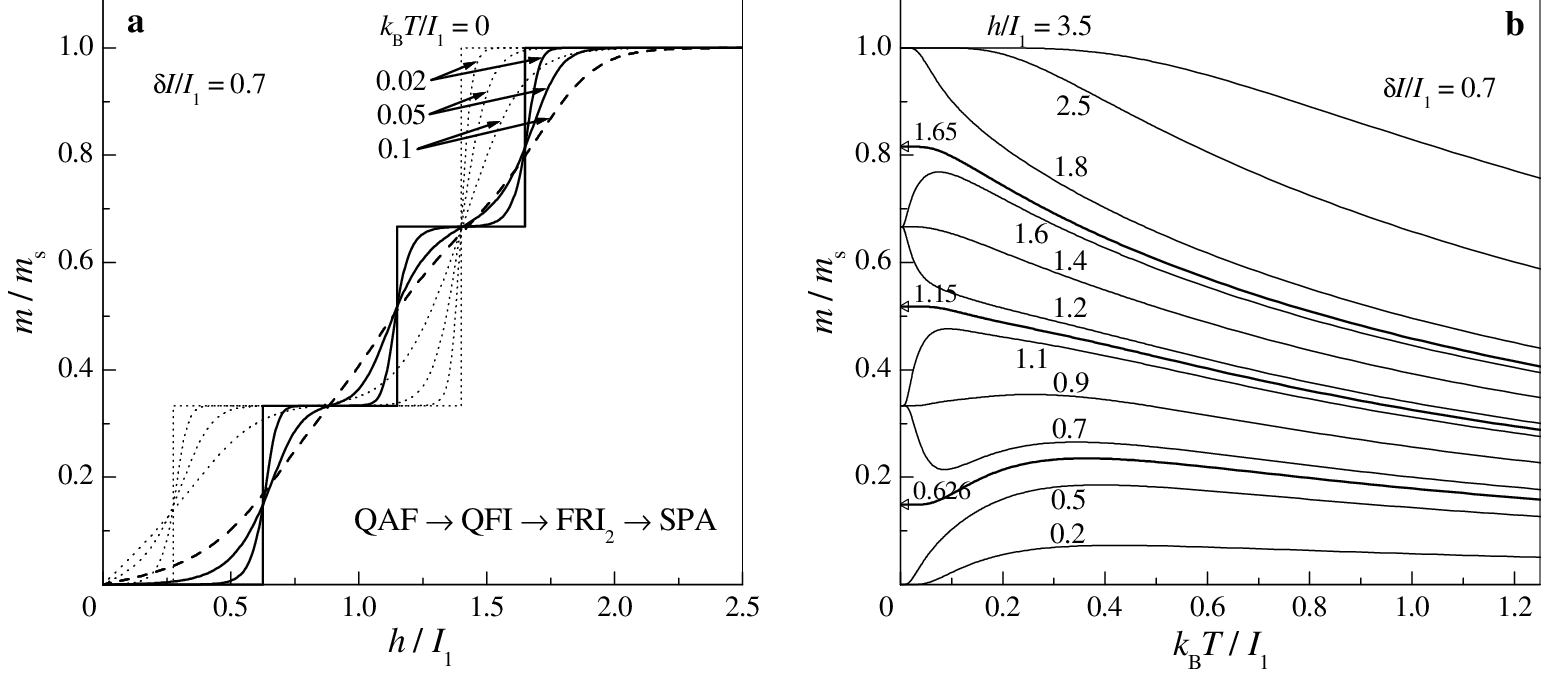}
\end{center}
\vskip-3mm
\caption{(a) The total magnetization as a function of the magnetic field at a few different temperatures for the particular case of the Heisenberg interaction $\Delta=1$ and $\tilde{J}=0.75$ by assuming one fixed value of the asymmetry parameter $\delta \tilde{I}=0.7$. The dotted lines correspond to the special case without the second-neighbor interaction $\tilde{I}_3=0$, the solid lines  to the particular case with the second-neighbor interaction $\tilde{I}_3=0.35$ of a moderate strength. (b) Thermal variations of the total magnetization for the particular case of the  Heisenberg interaction $\Delta=1$ and $\tilde{J}=0.75$, the asymmetry parameter $\delta \tilde{I}=0.7$ and the second-neighbor interaction $\tilde{I}_3=0.35$ at several values of the magnetic field. The triangle symbols denote critical fields at which two different ground states coexist together.}
\label{fig4}
\end{figure*}

Before proceeding to a discussion of finite-temperature properties, it is worth mentioning that the total magnetization of two ferrimagnetic ground states FRI$_1$ and QFI equals to one-third of the saturation magnetization in contrast to the total magnetization of the other ferrimagnetic ground state FRI$_2$ being equal to two-thirds of the saturation magnetization. For this reason, three remarkable ferrimagnetic ground states should manifest themselves in low-temperature magnetization curves as intermediate plateaux at one-third and/or two-thirds of the saturation magnetization. 
Let us consider first the field dependence of the total magnetization normalized with respect to the saturation magnetization as depicted in Figs.~\ref{fig3}(a) and \ref{fig4}(a) for a few temperatures and the isotropic Heisenberg coupling ($\Delta=1$) of the relative strength $\tilde{J}=0.75$. The dotted lines show the magnetization curves for the special case without the second-neighbor interaction ($\tilde{I}_3 = 0$), while the solid lines display the relevant change in the magnetization curves achieved upon switching on the second-neighbor interaction of moderate strength $\tilde{I}_3 = 0.35$. The most crucial change in the low-temperature magnetization curves due to the non-zero second-neighbor interaction $\tilde{I}_3$ definitely represents the novel two-thirds intermediate plateau connected with the ground state FRI$_2$. In fact, the two-thirds magnetization plateau may emerge both for low as well as high value of the asymmetry parameter as depicted in Figs.~\ref{fig3}(a) and \ref{fig4}(a) for two specific cases $\delta \tilde{I}=0.1$ and $0.7$, respectively, while the two-thirds plateau cannot be basically found in the magnetization process of the spin-1/2 Ising--Heisenberg diamond chain without the second-neighbor interaction \cite{lis3}. Moreover, it is quite obvious from Figs.~\ref{fig3}(a) and \ref{fig4}(a) that the asymmetry parameter $\delta \tilde{I}$ plays an essential role whether or not the magnetization curve might have plateau at zero magnetization, because the asymmetry parameter generally favors the QAF phase with a zero total magnetization before entering to the one-third plateau FRI$_1$ phase. The rising temperature generally smoothens the stepwise magnetization curves observable at low enough temperatures until the intermediate plateaux completely disappear from the magnetization curves.   

Next, let us turn our attention to temperature dependences of the total magnetization shown in Fig.~\ref{fig3}(b) and \ref{fig4}(b). The following general trends can be deduced from the displayed thermal variations of the total magnetization. The magnetization exhibits the marked temperature-induced changes whenever the external magnetic field is sufficiently close to critical fields determining a phase coexistence between two different ground states, whereas the vigorous thermally-induced increase (decrease) of the total magnetization is observable for the magnetic fields slightly below (above) the respective critical fields. Contrary to this, the magnetization falls down rather steadily with the rising temperature if the magnetic field is selected from the middle part of the magnetization plateau or above the saturation field. It is noteworthy that the monotonous decrease of the total magnetization with increasing temperature can be also found exactly at critical fields relevant to a phase coexistence of two different ground states, which are denoted in Fig.~\ref{fig3}(b) and \ref{fig4}(b) by triangle symbols. Under this condition, the total magnetization asymptotically reaches non-trivial values as temperature tends to zero, namely, $1/(3\sqrt{5}) \approx 0.1491$ for a coexistence point of the QAF-QFI phases, $(2 \sqrt{5} - 1)/(3\sqrt{5}) \approx 0.5176$ for a coexistence point of the FRI$_1$-FRI$_2$ and QFI-FRI$_2$ phases, $(2 \sqrt{5} + 1)/(3\sqrt{5}) \approx 0.7157$ for a coexistence point of the FRI$_2$-SPA phases.

\begin{figure}[t!]
\begin{center}
\includegraphics[width=0.95\columnwidth]{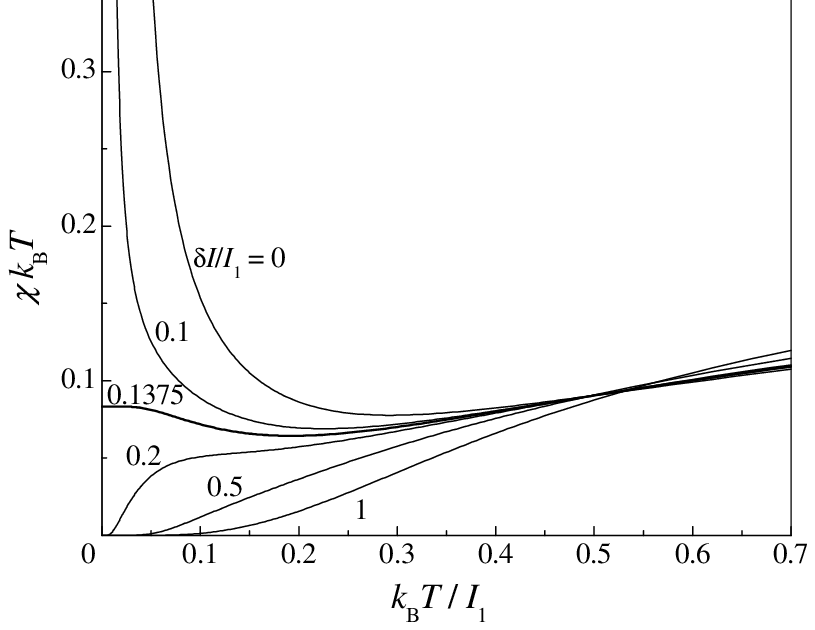}
\end{center}
\caption{The zero-field susceptibility times temperature product as a function of temperature for the particular case of the Heisenberg interaction $\Delta=1$ and $\tilde{J}=0.75$, 
the second-neighbor interaction $\tilde {I}_3=0.35$ and several values of the asymmetry parameter.}
\label{fig5}
\end{figure}

The temperature variation of the zero-field susceptibility times temperature product is depicted in Fig.~\ref{fig5} for the particular case of the isotropic Heisenberg coupling ($\Delta=1$, $\tilde{J}=0.75$) and the moderate strength of the second-neighbor interaction $\tilde {I}_3=0.35$. If the asymmetry parameter $\delta \tilde{I} < 0.1375$ is small enough in order to establish the ferrimagnetic ground-state FRI$_1$, the susceptibility times temperature product exhibits a striking non-monotonous dependence upon lowering temperature with a flat minimum preceding low-temperature divergence that is quite typical for ferrimagnets \cite{yam99}. On the other hand, the susceptibility times temperature product shows for higher values of the asymmetry parameter $\delta \tilde{I} > 0.1375$ a monotonous thermal dependence when it asymptotically tends to zero with decreasing temperature owing to the quantum antiferromagnetic ground state QAF. The stronger the antiferromagnetic second-neighbor interaction $\tilde {I}_3$ is, the less pronounced the temperature-induced increase of $\chi T$ product can be observed. 

\begin{figure}[h]
\begin{center}
\includegraphics[width=0.95\columnwidth]{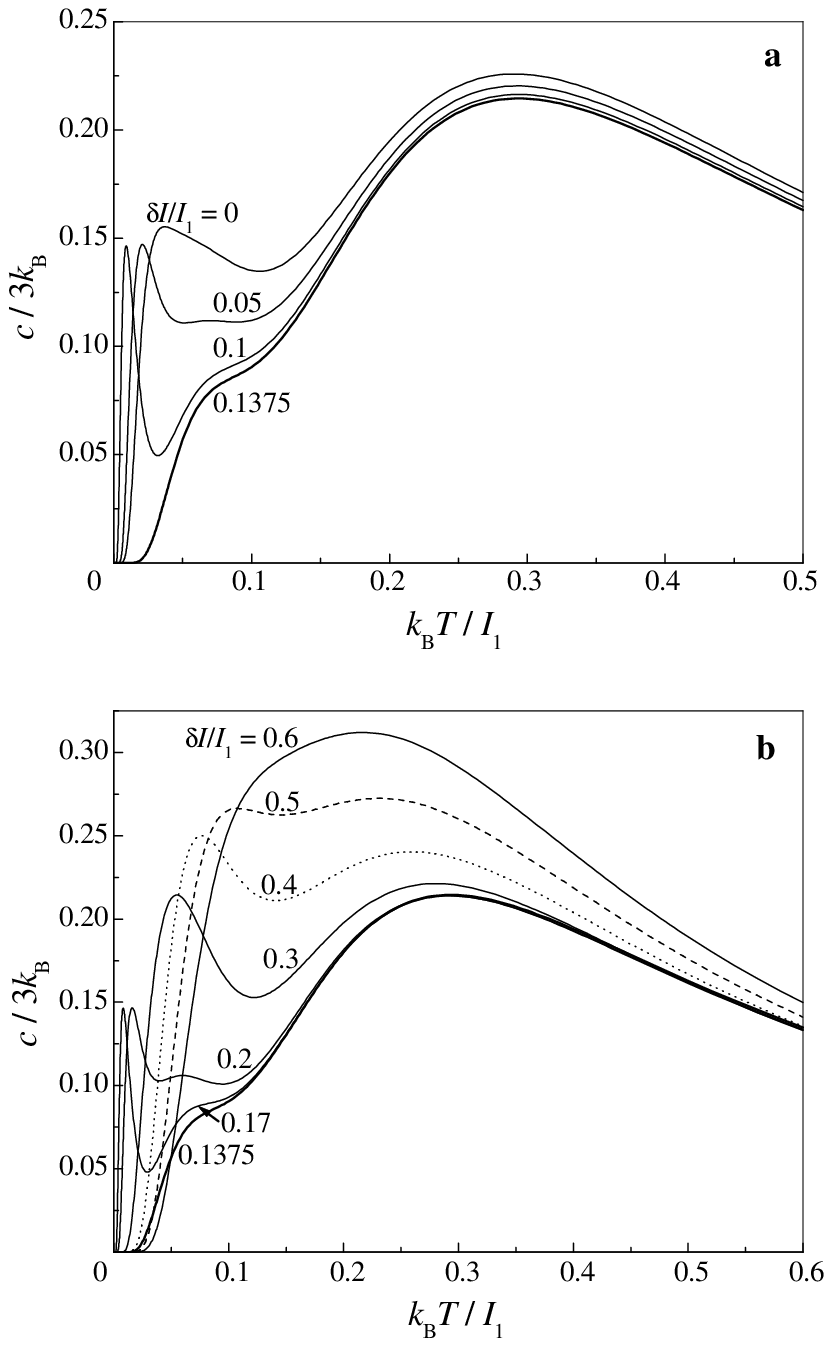}
\end{center}
\vskip-4mm
\caption{The temperature dependences of the zero-field specific heat for the Heisenberg interaction $\Delta=1$ and $\tilde{J}=0.75$, the second-neighbor interaction $\tilde {I}_3=0.35$ and several values of the asymmetry parameter $\delta \tilde{I}$. The particular values of the asymmetry parameter depicted in Fig. \ref{fig6}(a) coincide with the ground state FRI$_1$, while the ones 
displayed in Fig. \ref{fig6}(b) correspond to the ground state QAF.}
\label{fig6}
\end{figure}

Finally, let us examine in detail temperature variations of the zero-field specific heat. For this purpose, typical temperature dependences of the zero-field specific heat are plotted in Fig.~\ref{fig6} for the particular case of the isotropic Heisenberg coupling ($\Delta=1$, $\tilde{J}=0.75$), the second-neighbor interaction $\tilde {I}_3 = 0.35$ and several values of the asymmetry parameter $\delta \tilde{I}$. It can be clearly seen from Fig.~\ref{fig6}(a) that the round maximum observable at higher temperatures gradually decreases in height with increasing the asymmetry parameter $\delta \tilde{I}$ as far as the FRI$_1$ phase constitutes the ground state. Moreover, there also appears the additional Schottky-type peak at lower temperatures, which is shifted towards lower temperatures upon strengthening of the asymmetry parameter $\delta \tilde{I}$. The special case $\delta \tilde{I} = 0.1375$ corresponds to a phase coexistence between the FRI$_1$ and QAF phases, which is characterized through the notable thermal dependence of the heat capacity without the low-temperature peak but with a shoulder superimposed on ascending part of the round high-temperature maximum [see thick lines displayed in Fig.~\ref{fig6}(a) and (b)]. It worthwhile to remark that the double-peak temperature dependence of the specific heat is recovered if the asymmetry parameter $\delta \tilde{I}$ is strong enough to establish the QAF ground state [see Fig.~\ref{fig6}(b)]. In opposite to the previous case, the round high-temperature maximum increases in height with increasing the asymmetry parameter $\delta \tilde{I}$ and the low-temperature Schottky-type peak shifts towards higher temperatures until a complete coalescence of the low- and high-temperature peaks is achieved. The most spectacular thermal dependence of the heat capacity with three distinct round peaks can be detected for the asymmetry parameter close to $\delta \tilde{I} \approx 0.2$ when a mutual overlap of the low- and high-temperature peaks gives rise to a subtle intermediate (third) maximum significantly supported by the shoulder superimposed ascending part of the round high-temperature maximum [see the curve $\delta \tilde{I}=0.2$ in Fig.~\ref{fig6}(b)]. It is quite apparent that the sharpest peak observable at the lowest temperature can be always attributed to thermal excitations from the FRI$_1$ phase towards the QAF phase or vice versa. 

\section{Experimental implications}
\label{experiment}

In this section, let us draw a few implications for experimental representatives of the diamond spin chain Cu$_3$(CO$_3$)$_2$(OH)$_2$ (azurite) and the tetrahedral spin chain Cu$_3$Mo$_2$O$_9$ on the basis of the exactly solved spin-1/2 Ising--Heisenberg diamond chain with the second-neighbor interaction between the nodal spins. It has been argued in Refs. \cite{jes11,hon11} that the asymmetric spin-1/2 Heisenberg diamond chain with the second-neighbor interaction between the nodal spins provides a comprehensive description of all experimental data reported yet for the azurite. Although the generalized spin-1/2 Ising--Heisenberg diamond chain surely represents a considerable simplification of the analogous spin-1/2 Heisenberg diamond chain, it might be quite interesting to ascertain to what extent it explains the most pronounced features of the azurite because this simplified model still correctly reproduces the strongest Heisenberg interaction between the nearest-neighbor interstitial spins. According to Refs.~\cite{jes11,hon11}, the asymmetric spin-1/2 Heisenberg diamond chain with the second-neighbor interaction between the nodal spins quantitatively reproduces the experimental data of the azurite by assuming the following specific values of the exchange constants (see Fig.~\ref{fig1} for the notation used): $J/k_{\rm B}=33$~K, $I_1/k_{\rm B}=15.5$~K, $I_2/k_{\rm B}=6.9$~K, $I_3/k_{\rm B}=4.6$~K and the gyromagnetic ratio g = 2.06. This result actually implies that the Heisenberg coupling between the nearest-neighbor interstitial spins is by far the most dominant exchange interaction. 

\begin{figure}[h]
\begin{center}
\includegraphics[width=0.95\columnwidth]{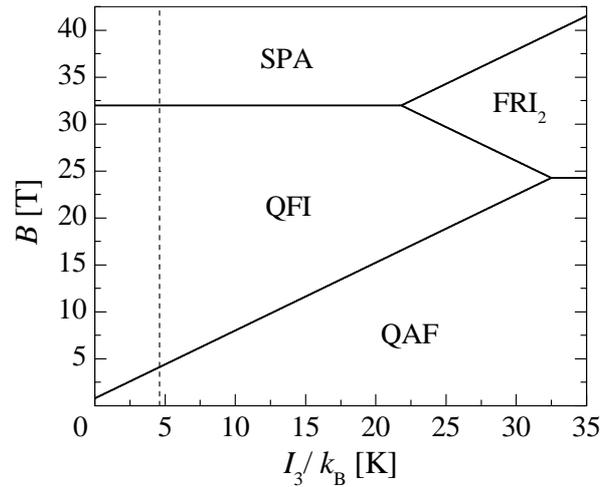}
\end{center}
\vskip-4mm
\caption{The ground-state phase diagram of the generalized spin-1/2 Ising--Heisenberg diamond chain in the $I_3 - B$ plane for the fixed values of exchange constants: $J/k_{\rm B}=33$~K ($\Delta=1$), $I_1/k_{\rm B}=15.5$~K, $I_2/k_{\rm B}=6.9$~K and the gyromagnetic ratio g = 2.06. The vertical broken line at $I_3/k_{\rm B}=4.6$~K shows the magnetization process relevant for the azurite.}
\label{fig7}
\end{figure}

The ground-state phase diagram of the generalized spin-1/2 Ising--Heisenberg diamond chain is depicted in Fig.~\ref{fig7} in the form of dependence magnetic field versus the second-neighbor interaction for the fixed values of exchange constants relevant for the azurite. The magnetization curve of the azurite should exhibit according to the generalized  Ising-Heisenberg diamond chain  an intermediate plateau at one-third of the saturation magnetization in the field range from 4.13~T to 31.96~T, which corresponds to the quantum ferrimagnetic phase QFI with a character of the dimer-monomer state. These results might be contrasted with the state-of-the-art DMRG data for the analogous spin-1/2 Heisenberg diamond chain, which predict the intermediate one-third plateau associated with the dimer-monomer state in the field range from $\simeq$9.5~T to $\simeq$31~T in accordance with the experimental magnetization data \cite{jes11,hon11}. While the lower edge of the one-third plateau is under-estimated within the generalized spin-1/2 Ising--Heisenberg diamond chain approximatively two times, the upper edge of the one-third plateau quantitatively coincides almost exactly with the experimental data and the relevant results of the generalized spin-1/2 Heisenberg diamond chain (the estimated error is around 3~\%). Altogether, it could be concluded that the generalized spin-1/2 Ising--Heisenberg diamond chain not only qualitatively reproduces a character of the ferrimagnetic dimer-monomer state within the one-third plateau region, but it quantitatively reproduces the upper edge of the one-third magnetization plateau. The main reason for this surprisingly good quantitative concordance is the fact that the quantum ($xy$) part of the exchange interactions $I_1$, $I_2$ and $I_3$ becomes irrelevant once the nodal spins are fully polarized by the magnetic field within the one-third magnetization plateau corresponding to the quantum dimer-monomer state. It is quite tempting to conjecture, moreover, that the two-thirds plateau could be detected in the magnetization curve of the azurite just if the second-neighbor interaction between the nodal spins would be greater than $I_3/k_{\rm B} \geq 21.8$~K, i.e., if it would be roughly five times stronger than it is in reality. 
  
Last but not least, let us employ the exact solution for the generalized spin-1/2 Ising--Heisenberg diamond chain to gain some insight into the magnetism of the copper-based chain of corner-sharing tetrahedra Cu$_3$Mo$_2$O$_9$ \cite{hase08,hama08,kuro11,hase11,mats12,kuro14} to be further referred to as the distorted tetrahedral chain. First, it is worth mentioning that the asymmetric spin-1/2 Ising--Heisenberg diamond chain accounting for the additional second-neighbor interaction between the nodal spins is isomorphous with the distorted spin-1/2 Ising--Heisenberg tetrahedral chain with four different exchange interactions within the tetrahedron unit. Even though the magnetic compound Cu$_3$Mo$_2$O$_9$ is again the experimental realization of the distorted spin-1/2 Heisenberg tetrahedral chain, it is our hope that the simplified spin-1/2 Ising--Heisenberg tetrahedral chain may capture some important vestiges of its magnetic behavior. Recent experimental measurements performed on the distorted tetrahedral chain Cu$_3$Mo$_2$O$_9$ serve in evidence of the spectacular quantum antiferromagnetic order, in which the uniform N\'eel order of the nodal spins along the chain direction is accompanied with the spin-singlet-like state of the interstitial spins \cite{hase08,hama08,kuro11,hase11,mats12,kuro14}. According to our notation (see Fig.~\ref{fig1}), the following exchange constants have been extracted from the inelastic neutron scattering data for two strongest exchange interactions $J/k_{\rm B}=67$~K, $I_3/k_{\rm B}=75$~K, and the respective difference between two weaker exchange interactions $\delta I/k_B = (I_1 - I_2)/k_{\rm B} = 35$~K \cite{mats12}. Despite the fact that the absolute values of two weaker interactions $I_1$ and $I_2$ cannot be simply figured out from the available experimental data and they are still under debate, the rough estimate of the weakest interaction is around $I_2/k_{\rm B} \approx 12$~K \cite{kuro11}. Regardless of the aforementioned ambiguity, the strongest exchange interaction $I_3/k_{\rm B}=75$~K definitely drives the zero-field ground state into the unique quantum antiferromagnetic phase (\ref{GS})-(\ref{PA}) with the N\'eel order of the nodal spins and the spin-singlet-like state of the interstitial spins characterized by the staggered magnetization:
\begin{eqnarray}
m_{\rm stag} &=&  \langle {\rm QAF}| \frac{1}{2} (\hat{S}_{k,1}^z - \hat{S}_{k,2}^z) | {\rm QAF} \rangle  \nonumber \\
&=& \frac{1}{2} \frac{\delta I}{\sqrt{(\delta I)^2 + (J \Delta)^2}} \simeq 0.23,
\label{sm}
\end{eqnarray}
\begin{figure}[h]
\begin{center}
\includegraphics[width=0.95\columnwidth]{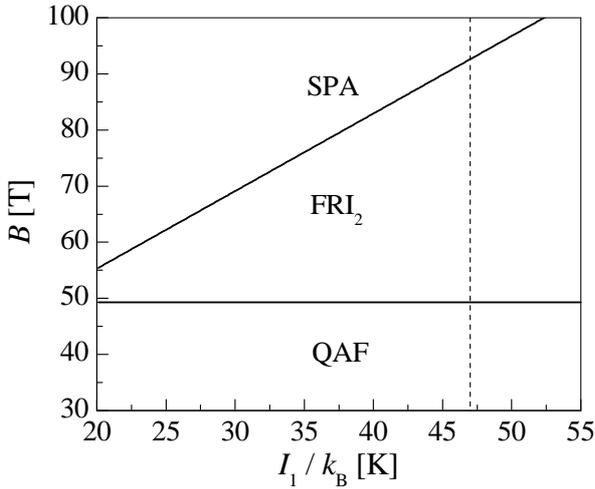}
\end{center}
\vskip-4mm
\caption{The ground-state phase diagram of the generalized spin-1/2 Ising--Heisenberg diamond chain in the $I_1 - B$ plane for the fixed values of exchange constants: $J/k_{\rm B}=67$~K ($\Delta=1$), $I_3/k_{\rm B}=75$~K, $\delta I/k_B = (I_1 - I_2)/k_{\rm B} = 35$~K and the gyromagnetic ratio g = 2.154 that are relevant for the distorted tetrahedral chain Cu$_3$Mo$_2$O$_9$. The vertical broken line at $I_1/k_{\rm B}=47$~K shows the estimated magnetization process.}
\label{fig8}
\end{figure}
which implies a quantum reduction of the magnetic moment of interstitial spins roughly to 47~\% of its saturation value. It is worth noticing that the quantum reduction of staggered magnetization depends just on a relative strength of the coupling $J$ between the nearest-neighbor interstitial spins and the difference of exchange interactions $\delta I = I_1 - I_2$, which are known quite accurately from the experimental data unlike the absolute values of the exchange interactions $I_1$ and $I_2$. With this background, we have constructed for the distorted tetrahedral chain Cu$_3$Mo$_2$O$_9$ the ground-state phase diagram in the $I_1-B$ plane displayed in Fig.~\ref{fig8}. The interaction constants estimated for the distorted tetrahedral chain Cu$_3$Mo$_2$O$_9$ evidently fall into the parameter region, where the intermediate one-third magnetization plateau is absent but there exists the two-thirds magnetization plateau connected to the classical FRI$_2$ ground state emerging at sufficiently high magnetic fields. This theoretical prediction is consistent with recent high-field measuremens performed on the single-crystal sample of Cu$_3$Mo$_2$O$_9$, which give a clear evidence for the two-thirds magnetization plateau the microscopic origin of which is currently under investigation \cite{kuro14}. It is worthwhile to remark that the lower edge of indermediate two-thirds plateau is independent of the absolute value of the exchange constant $I_1$ (it depends only on the exchange constant $J$ and the difference $\delta I$), which allows us to fix the lower edge of two-thirds plateau quite accurately to the value $B = 49.3$~T that is in a relatively good quantitative accord with the values $B = 52.3$, $60.3$ and $47.5$~T reported for the magnetization data measured along three crystalographic axes of the distorted tetrahedral chain Cu$_3$Mo$_2$O$_9$ \cite{kuro14}. From this perspective, one may infer that the two-thirds plateau actually bears a connection with the classical ferrimagnetic ground state FRI$_2$.

\section{Conclusion}
\label{conclusion}

In the present article, the ground state and thermodynamics of the asymmetric spin-1/2 Ising--Heisenberg diamond chain generalized by the second-neighbor interaction between the nodal spins are examined by a rigorous calculation. Exact results for the free energy, magnetization, susceptibility, entropy and heat capacity of the generalized spin-1/2 Ising--Heisenberg diamond chain have been derived by applying the method of decoration-iteration transformation. In particular, our attention has been focused on exploring the magnetic behavior of the generalized spin-1/2 Ising--Heisenberg diamond chain with the antiferromagnetic interactions, which should exhibit the most intriguing magnetic features in relation with a strong interplay between the geometric frustration and local quantum fluctuations. 

Among the most interesting results one could mention a considerable diversity of ground-state phase diagrams, which may include in total five different ground states: the saturated paramagnetic ground state SPA, two classical ferrimagnetic ground states FRI$_1$ and FRI$_2$, one quantum ferrimagnetic ground state QFI and the unique quantum antiferromagnetic ground state QAF. Notably all ferrimagnetic ground states should manifest themselves in low-temperature magnetization curves as intermediate plateaux at fractional values of the saturation magnetization. While the total magnetization of two translationally invariant classical and quantum ferrimagnetic phases FRI$_1$ and QFI equals to one-third of the saturation magnetization, the total magnetization of the other classical ferrimagnetic phase FRI$_2$ (up-up-up-down-up-up-...) with a translationally broken symmetry equals to two-thirds of the saturation magnetization. It is worthy of notice that the peculiar two-thirds magnetization plateau related to the classical ferrimagnetic phase FRI$_2$ cannot be definitely found in the spin-1/2 Ising--Heisenberg diamond chain without the second-neighbor interaction between the nodal spins \cite{lis3} but the four-spin coupling might represent an alternative mechanism for a stabilization of the two-thirds plateau \cite{gal13,gal14}. Besides, we have also demonstrated a rich variety of temperature dependences of the zero-field susceptibility and zero-field specific heat, whereas thermal dependences of zero-field specific heat may display one or two anomalous low-temperature peaks in addition to the round maximum observable at higher temperatures.   

The exact solution presented for the generalized spin-1/2 Ising--Heisenberg diamond chain has also proved its usefulness in elucidating magnetic properties of two copper-based chains Cu$_3$(CO$_3$)$_2$(OH)$_2$ and Cu$_3$Mo$_2$O$_9$, which provide outstanding experimental realizations of the diamond spin chain and the distorted tetrahedral spin chain, respectively.
As a matter of fact, the generalized spin-1/2 Ising--Heisenberg diamond chain correctly reproduces the intermediate one-third magnetization plateau of the azurite as macroscopic manifestation of the quantum ferrimagnetic (dimer-monomer) phase, whereas an upper edge of the intermediate plateau coincides almost exactly with the experimental results and the state-of-the-art numerical calculations for the analogous but more sophisticated Heisenberg model \cite{jes11,hon11}. Moreover, the exactly solved spin-1/2 Ising--Heisenberg diamond chain sheds light on the spectacular quantum antiferromagnetic state QAF of the distorted tetrahedral chain Cu$_3$Mo$_2$O$_9$, which is characterized by the N\'eel order of the nodal spins and the spin-singlet-like state of the interstitial spins. Our rigorous results have enabled us to conjecture to what extent the staggered magnetization of interstitial spins is reduced by quantum fluctuations within the QAF, 
as well as, to propose the microscopic nature of two-thirds magnetization plateau verified by recent high-field measurements \cite{kuro14}.            

\begin{acknowledgement}
B.L. acknowledges the financial support provided by the National Scholarship Programme of the Slovak Republic for the Support of Mobility of Students, PhD Students, University Teachers,
Researchers and Artists. J.S. acknowledges the financial support provided by the grant of The Ministry of Education, Science, Research and Sport of the Slovak Republic under the contract No.
VEGA 1/0234/12 and by the grants of the Slovak Research and Development Agency under the contracts Nos. APVV-0132-11 and APVV-0097-12.
\end{acknowledgement}

\end{document}